\documentclass[10pt,letterpaper]{article}

\usepackage[T1]{fontenc}
\usepackage[utf8]{inputenc}
\usepackage{newtxtext,newtxmath}
\usepackage{microtype}
\usepackage[letterpaper,top=0.68in,bottom=0.72in,left=0.74in,right=0.74in,
            headheight=15pt,headsep=16pt,footskip=26pt]{geometry}
\usepackage{graphicx}
\usepackage{xcolor}
\usepackage{tikz}
\usetikzlibrary{arrows.meta,positioning,calc}
\usepackage{booktabs}
\usepackage{float}
\usepackage{array}
\usepackage{tabularx}
\usepackage{needspace}
\usepackage{enumitem}
\usepackage{amsmath}
\usepackage{titlesec}
\usepackage{fancyhdr}
\usepackage[font=small,labelfont={bf,rm},labelsep=period]{caption}
\usepackage[numbers,square,sort&compress]{natbib}
\usepackage[most]{tcolorbox}
\usepackage{hyperref}

\definecolor{Cobalt}{HTML}{1951E5}
\definecolor{Teal}{HTML}{39BCAE}
\definecolor{Ink}{HTML}{192238}
\definecolor{Subtle}{HTML}{637084}
\definecolor{SoftRule}{HTML}{DDE4EF}
\definecolor{Pale}{HTML}{F5F8FC}
\definecolor{PaleBlue}{HTML}{EDF3FF}
\hypersetup{colorlinks=true,linkcolor=Cobalt,citecolor=Cobalt,urlcolor=Cobalt,
  pdftitle={Mach2: Domain-Specific Agent Packages and Endpoint-Bound Execution},
  pdfauthor={C-PRIME}}

\newcommand{\papertitle}{Mach2: Scientific Discovery Platform with \\Deterministic Execution Constraints}
\newcommand{\papersubtitle}{Model-authored orchestration, portable engineering agents, and result reconciliation across execution boundaries}

\newcommand{\paperauthors}{%
  Vansh Sharma\textsuperscript{1,2,*},
  Venkat Raman\textsuperscript{1,2}}
\newcommand{\paperdate}{November 2026}
\newcommand{\paperaffiliations}{%
  \textsuperscript{1}Center for Prediction, Reasoning \& Intelligence for Multiphysics Exploration (C-PRIME)\\[2pt]
  \textsuperscript{2}Department of Aerospace Engineering, University of Michigan}

\newcommand{\papercorrespondence}{Vansh Sharma (\href{mailto:vanshs@umich.edu}{vanshs@umich.edu})}

\color{Ink}
\setlist[itemize]{topsep=3pt,itemsep=3pt,parsep=0pt,leftmargin=1.35em,
  label={\textcolor{Cobalt}{\large\textbullet}}}
\titleformat{\section}{\rmfamily\bfseries\large\color{Ink}}
  {\textcolor{Cobalt}{\thesection}}{0.70em}{}
\titleformat{\subsection}{\rmfamily\bfseries\normalsize\color{Ink}}
  {\textcolor{Cobalt}{\thesubsection}}{0.70em}{}
\titlespacing*{\section}{0pt}{17pt plus 2pt minus 1pt}{6pt}
\titlespacing*{\subsection}{0pt}{11pt plus 1pt minus 1pt}{4pt}
\fancypagestyle{cprime}{%
  \fancyhf{}%
  \fancyhead[L]{\rmfamily\fontsize{8}{10}\selectfont\bfseries\color{Ink}C-PRIME}%
  \fancyhead[R]{\rmfamily\fontsize{7.5}{9}\selectfont\color{Subtle} MACH2 Intelligence System}%
  \fancyfoot[R]{\rmfamily\fontsize{8.5}{10}\selectfont\bfseries\color{Cobalt}\thepage}%
}
\fancypagestyle{cprimefirst}{%
  \fancyhf{}%
  \fancyfoot[L]{\rmfamily\fontsize{7.6}{9.2}\selectfont\color{Subtle}\textcopyright\ 2026 All rights reserved by the Author and C-PRIME.}%
  \fancyfoot[R]{\rmfamily\fontsize{8.5}{10}\selectfont\bfseries\color{Cobalt}\thepage}%
}
\newtcolorbox{abstractpanel}{
  enhanced,breakable,boxrule=0pt,colback=Pale,
  borderline west={2.2pt}{0pt}{Cobalt},
  sharp corners,left=13pt,right=12pt,top=11pt,bottom=10pt,
  before skip=0pt,after skip=8pt
}
\newcommand{\minilabel}[1]{{\rmfamily\fontsize{8.5}{10}\selectfont\bfseries\color{Cobalt}\MakeUppercase{#1}}}

\begin{document}
\thispagestyle{cprimefirst}

% --- C-PRIME masthead: compact original logo, width 1.72 in ---
\noindent\begin{minipage}[c]{0.60\linewidth}
  \includegraphics[width=1.72in]{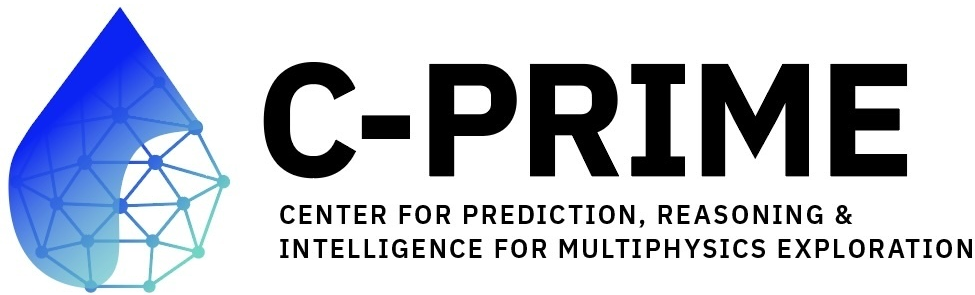}
\end{minipage}\hfill\begin{minipage}[c]{0.37\linewidth}
  \raggedleft\rmfamily\fontsize{7.6}{10.2}\selectfont
  \color{Subtle}RESEARCH SERIES\quad$\bullet$\quad \paperdate
\end{minipage}

\vspace{2pt}
\noindent\textcolor{Cobalt}{\rule{0.62in}{1.6pt}}%
\textcolor{SoftRule}{\rule{\dimexpr\linewidth-0.62in\relax}{0.8pt}}
\vspace{10pt}

{\rmfamily\bfseries\fontsize{23}{27}\selectfont\color{Ink}\papertitle\par}
\vspace{5pt}
{\rmfamily\fontsize{9.7}{12.4}\selectfont\color{Subtle}\papersubtitle\par}
\vspace{12pt}
{\rmfamily\fontsize{9.9}{12.4}\selectfont\bfseries\color{Ink}\paperauthors\par}
\vspace{4pt}
{\rmfamily\fontsize{8.9}{12.2}\selectfont\color{Subtle}\paperaffiliations\par}
\vspace{12pt}

\begin{abstractpanel}
  \minilabel{Abstract}\par\vspace{3pt}
Language-model agents can formulate workflows and invoke specialized tools, yet compliance with natural-language operating rules is not guaranteed by prompting alone. Mach2 addresses this limitation by coupling Eigen orchestration to deterministic, lint-like checks of tool requests, source changes, acceptance tests, and artifact contracts. Domain requirements are represented as executable predicates: mandatory failures prevent advancement of the checked operation, while diagnostic findings are retained for correction or review. User-specified test-driven development (TDD), CodeGrapher's abstract-syntax-tree analysis, and history-guided evolutionary search provide complementary verification and optimization mechanisms. Mach2 also generates task-specific agent packages for execution through independently selected endpoints. The originating harness validates returned artifacts against declared interfaces and evidence requirements without assuming continuous supervision of the remote run. Model-internal-knowledge (MIK) dependencies are recorded separately from source-grounded and test-supported findings. This paper formalizes the rule-gated execution model and the agent-package lifecycle, and proposes measurable criteria for evaluating rule compliance, engineering validity, and cross-endpoint task continuity.

\newcommand{\paperkeywords}{Agentic Harness; Multi-agent Orchestration; Domain-specific Harness; Secure Agentic Platform; Detached Execution; Verification}

  % The two metadata lines below belong INSIDE the abstract panel.
  \par\vspace{8pt}
  {\color{SoftRule}\rule{\linewidth}{0.55pt}}\par\vspace{2pt}
  {\rmfamily\fontsize{8.5}{11.5}\selectfont\color{Subtle}%
    \textbf{\textcolor{Ink}{Keywords}}\quad\paperkeywords\par
    \vspace{3pt}
    \textbf{\textcolor{Ink}{Corresponding author}}\quad\papercorrespondence\par
  }
\end{abstractpanel}

\section{Introduction}
Language-model agents extend generative models with tools, persistent state, and delegated workflows. Written instructions can define intended behavior, but they do not ensure that each generated tool call, code modification, or final artifact satisfies the specified constraints. Agent and their underlying models tend to defy written instructions. Evaluations of verifiable instruction following and tool-using agents document deviations from stated requirements, including failures induced by ``untrusted" tool output~\cite{zhou2023ifeval,debenedetti2024agentdojo}. For computational engineering, a rule violation may invalidate a result even when an implementation appears to complete successfully. The objective of this report is to introduce a user-aligned agentic development framework that integrates explicit governance mechanisms, enforceable execution constraints, and verifiable workflows to support reliable and accountable autonomous operation.

Mach2 is a scientific discovery platform (SDP) built on an OpenCode-derived runtime. It combines Eigen's (a custom built agent) model-authored orchestration with user-specified tests, structural code analysis, and executable engineering tools. In contrast to the original MachOne system developed in Sharma's doctoral research~\cite{sharma2026thesis}, Mach2 prioritizes the deep integration of scientific tools and the management of long-running workflows, with scoped task context and provenance checks designed to mitigate context contamination and maintain traceability of scientific outputs. OpenCode supports configurable agents, plugins, and server interfaces~\cite{opencodeagents,opencodeplugins,opencodeserver}; DeepSeek Harness also supports model-authored workflows and subagent coordination~\cite{dshworkflow}. The distinguishing architectural emphasis here is the coupling of these capabilities to \emph{runtime-checkable constraints} and to agent packages that preserve task contracts across execution endpoints.

The resulting lifecycle has three verifiable boundaries: compliance checks around local actions, acceptance checks on candidate engineering artifacts, and contract validation when external execution returns. Mach2 also incorporates test-driven scientific-code construction following Chain of Unit-Physics~\cite{sharma2025unitphysics} and history-guided optimization based on AUTO~\cite{carreon2025auto}.

\section{Mach2 Orchestration and Engineering Runtime}
\subsection{Eigen coordination and user-defined acceptance}
Eigen agent decomposes user objectives into bounded tasks, generates execution workflows, commissions specialist agents, and integrates results. In the user-facing test-driven development (TDD) interaction, expected behavior and acceptance tests are specified before implementation. A baseline execution records unmet requirements; subsequent revisions are accepted only after the applicable tests are re-executed. For scientific code, tests can express conservation, dimensional consistency, and physical limiting behavior, consistent with Chain of Unit-Physics~\cite{sharma2025unitphysics}.

CodeGrapher indexes source-code entities and relationships obtained through language-aware abstract syntax trees (ASTs). The patent application by Sharma and Raman describes syntax-tree traversal, graph-backed source retrieval, and verification of generated assertions against source locations~\cite{sharma2026falseinformation}. CADGrapher integrates supported geometry operations through FreeCAD. These services supply structured evidence and executable actions to Eigen rather than relying exclusively on model-generated descriptions.

\subsection{Workspace-scoped execution}
Mach2 is distributed in a Linux ARM64 Docker runtime. Its launcher mounts the selected host project as the working directory and preserves project outputs there; application state persists in a separate volume. Host filesystem exposure is determined by the configured mounts. Internal temporary storage can remain available, so the launch-workspace model concerns accessible host files and durable artifacts rather than an unconditional prohibition on temporary paths.

\begin{table}[H]
\caption{Mach2 execution components and associated checks.}
\label{tab:components}
\centering
\rmfamily\fontsize{8.8}{11.7}\selectfont
\renewcommand{\arraystretch}{1.23}
\begin{tabularx}{\linewidth}{@{}p{2.03in}>{\raggedright\arraybackslash}X@{}}
\toprule
\textbf{Component} & \textbf{Execution responsibility}\\
\midrule
Eigen & Dependency-aware task planning, branch-local agent proposals, and integration\\
User interaction (TDD) & Testable requirements, baseline failure, and acceptance evidence\\
Deterministic rule checks & Pre-merge and post-merge acceptance gates, with diagnostics\\
CodeGrapher / CADGrapher & AST-grounded code queries and executable CAD operations\\
Agent-package generator & Creates specialized agents, endpoint interfaces, and contracts\\
Result reconciliation & Checks returned outputs and resumes the originating task\\
\bottomrule
\end{tabularx}
\end{table}

\section{Deterministic Enforcement of Written Rules}
\subsection{From instructions to executable predicates}
Natural-language requirements are operational specifications, not enforcement mechanisms. In Mach2, eligible rules are translated into checkable predicates resembling static-analysis or lint rules. Their subjects include permitted tools and filesystem paths, allowable source changes, required test evidence, output schemas, and provenance records. This separates a model's choice of an action from the harness's decision to permit or accept it.

Let $x$ denote the available task state, proposed operation, and associated artifacts, and let $\rho_j(x)$ be the result of rule $j$:
\begin{equation}
\rho_j(x)\in\{\mathrm{pass},\mathrm{fail},\mathrm{review}\},\qquad
G(x)=\bigwedge_{j\in\mathcal H}[\rho_j(x)=\mathrm{pass}],
\label{eq:rulegate}
\end{equation}
where $\mathcal H$ contains mandatory rules. A gated action or artifact can advance only when $G(x)$ is true. Nonmandatory findings generate diagnostics, while an unresolvable mandatory check requires correction or an explicitly authorized change to the specification. Evaluation results are retained with rule identifiers and the corresponding task or artifact revisions.

\subsection{Rule checkpoints and scope of assurance}
Checks are applied at the appropriate lifecycle boundary: before an instrumented tool action (permissions and target paths), after a code modification (AST invariants and TDD outcomes), during package construction (interface and resource constraints), and upon result import (required fields, task identity, and available scientific evidence). An agent's textual declaration of success cannot substitute for a passing rule evaluation. Determinism applies to the \emph{implemented predicates on observed inputs}; it does not establish that a natural-language requirement was translated perfectly, that the test suite is complete, or that an unobserved external endpoint complied with local restrictions.

\subsection{Orchestrated execution and state acceptance}

\begin{figure}[h]
    \centering
    \includegraphics[width=0.75\linewidth]{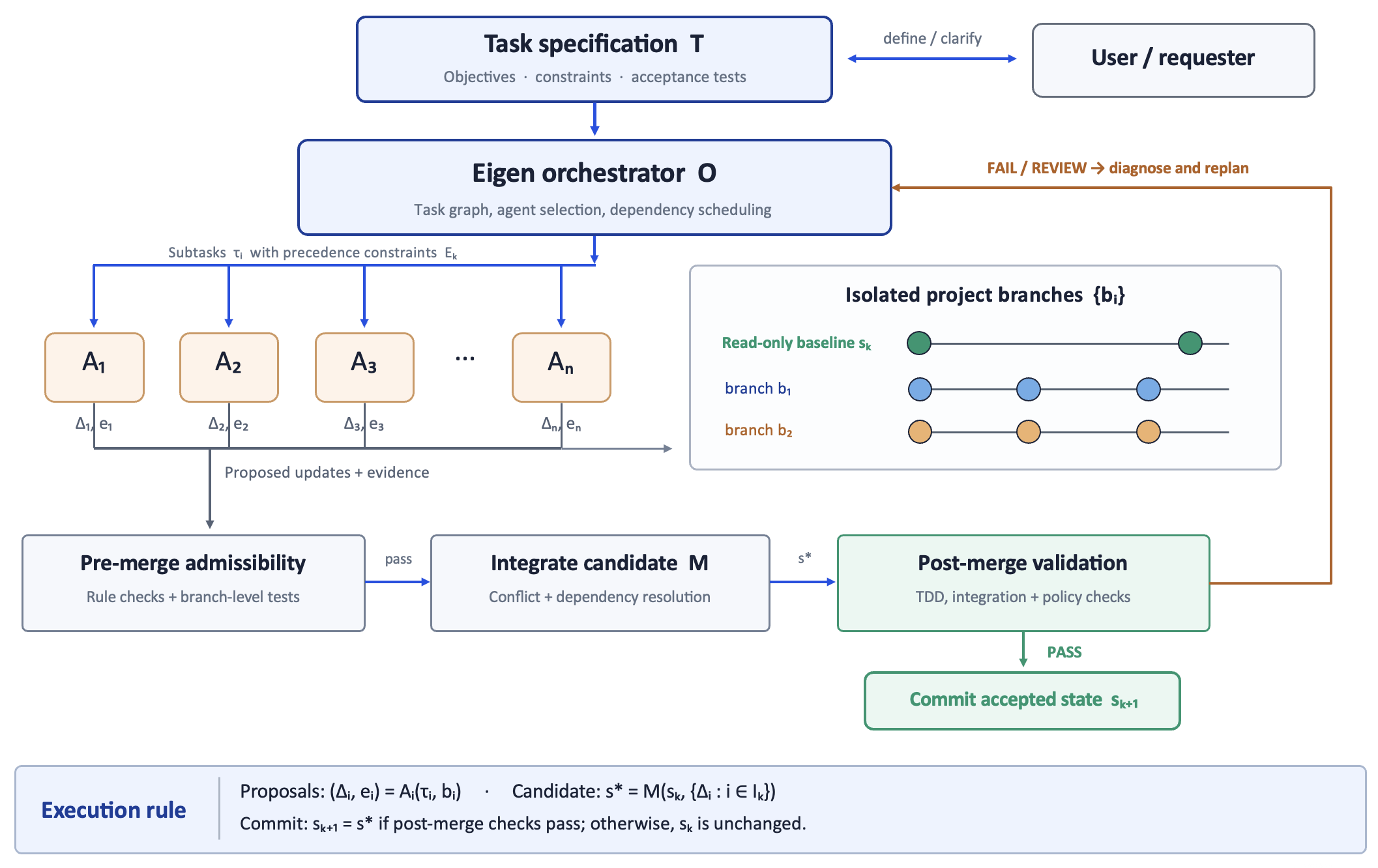}
    %\vspace{-7mm}
    \caption{Rule-gated multi-agent execution. Eigen decomposes the task into
dependent subtasks, receives branch-local proposals and evidence, and integrates
admissible changes into a candidate state. Only a passing post-merge validation
updates the accepted project state; failed checks trigger replanning.}
    \label{fig:execution}
\end{figure}

The orchestrator treats the accepted project state $s_k$ as a fixed baseline for an execution cycle. From a task specification $T$, Eigen constructs a directed acyclic task graph $\mathcal D_k=(\mathcal T_k,E_k)$, where $E_k$ denotes precedence constraints, and assigns the $N$ currently runnable subtasks $\tau_i\in\mathcal T_k$ to workers $A_i$ (see Fig.~\ref{fig:execution}). Each worker operates on a logically isolated branch $b_i$ derived from $s_k$ and returns a proposed modification $\Delta_i$ with supporting evidence $e_i$:
\begin{equation}
\mathcal O(T,s_k)\leadsto \mathcal D_k,\qquad
(\Delta_i,e_i)=A_i(\tau_i,b_i),\quad b_i=B_i(s_k).
\label{eq:agentproposals}
\end{equation}
Proposals are screened by the applicable mandatory predicates in Eq.~\eqref{eq:rulegate}. For $x_i=(s_k,\tau_i,\Delta_i,e_i)$, the integration set $\mathcal I_k$ contains only proposals passing the applicable mandatory rules, satisfying their dependencies, and remaining free of unresolved change conflicts. Integration produces a \emph{candidate} state $s^\star_{k+1}$, not an immediate update to the accepted project:
\begin{equation}
\begin{aligned}
\mathcal I_k&\subseteq\bigl\{i: G(x_i)=1\bigr\},\\
s^\star_{k+1}&=M\bigl(s_k,\{\Delta_i:i\in\mathcal I_k\}\bigr),\\
s_{k+1}&=\begin{cases}
s^\star_{k+1},&V_{\mathrm{post}}(s^\star_{k+1};T)=\mathrm{pass},\\
s_k,&\text{otherwise}.
\end{cases}
\end{aligned}
\label{eq:accepttransition}
\end{equation}
Here $M$ is the conflict-aware integration operation and $V_{\mathrm{post}}$ comprises integration tests and project-level acceptance rules. A failed or review-required check leaves the accepted state unchanged and returns its diagnostic evidence to Eigen for corrective work or replanning. Branches may be realized as version-control branches or equivalent isolated workspaces; the state-transition model does not require a particular version-control implementation.

\section{Domain-Specific Agent Package Generation}
\subsection{Rule-bearing harness construction}
A domain specification comprises objectives, permissible tools, evidence, policies, verifiers, interfaces, and resource limits:
\begin{equation}
D=(\mathcal O,\mathcal T,\mathcal K,\mathcal P,\mathcal V,\mathcal I,\mathcal B),
\quad H_D=\mathcal C(D;R,M)=(A,\Pi,\Gamma,\Sigma,\mathcal I).
\label{eq:compilation}
\end{equation}
Here $\mathcal P$ contains the rule predicates and their severity, $\mathcal V$ the test and domain acceptance conditions, $A$ the agent roles, $\Pi$ the execution policy, $\Gamma$ the bound tools and sources, and $\Sigma$ the persistent task record. The compiler $\mathcal C$ creates an executable agent configuration for compatible runtime $R$ and model service $M$. Compilation means constructing and binding a workflow and its checks, not compiling model weights.

\subsection{Portable agents and interface contracts}
For independent execution, the generated artifact is $P_D=(H_D,E_D,Q_D)$, where $E_D$ defines the endpoint requirements and $Q_D$ the returned data and evidence schema. Mach2 can create specialized agent files for tasks that cannot remain under Eigen's supervision; the user supplies a compatible endpoint and initiates execution. Required identifiers, artifact formats, and applicable acceptance rules accompany the package. Credentials and model weights need not be embedded in it. Local rules can govern export and subsequent acceptance; guarantees about the intervening remote run depend on controls present at that endpoint.

\section{Test-Driven and Evolutionary Execution}
For supervised development, Eigen records the current state $s_k$, action $a_k$, and observation $o_k$ at each iteration:
\begin{equation}
(\widetilde{s}_{k+1},o_k)=F_{H_D}(s_k,a_k),\qquad
v_k=V_D(\widetilde{s}_{k+1},o_k).
\end{equation}
The tentative state $\widetilde{s}_{k+1}$ is evaluated before commitment. The verifier $V_D$ includes user-defined TDD tests and applicable domain checks; the transition to an accepted state follows Eq.~\eqref{eq:accepttransition}. Bounded retries can respond to failing tests, while the mandatory rule gate in Eq.~\eqref{eq:rulegate} independently governs permitted actions and acceptance. Persistent agent iteration has precedents in the Ralph workflow~\cite{huntleyralph,anthropicralph}; here the completion condition is external test evidence rather than the model's own completion statement.

For design exploration, the AUTO methodology employs evaluated candidate history to select exploration, recombination, or refinement~\cite{carreon2025auto}. Rule and test checks establish admissibility; objective measurements rank candidates that satisfy those constraints. This separates \emph{feasibility enforcement} from \emph{evolutionary search} and allows failed candidates to inform subsequent trials without being accepted as valid engineering outputs.

\section{External Agent Execution and Reconciliation}
\begin{figure}[H]
\centering
\begin{tikzpicture}[node distance=3.0mm and 6.0mm,>=Latex,
 every node/.style={font=\footnotesize},
 block/.style={draw=SoftRule,rounded corners=2pt,fill=Pale,align=center,minimum height=0.81cm,text width=3.15cm,inner sep=3.9pt},
 macharrow/.style={-Latex,thick,draw=Cobalt},
 feedback/.style={-Latex,thick,draw=Subtle,dashed}]
\node[block] (task) {User specification\\tests / rules};
\node[block,right=of task] (eigen) {Eigen orchestration\\rule-gated tool actions};
\node[block,right=of eigen] (package) {Agent package\\policy / result contract};
\node[block,right=of package] (endpoint) {User-linked endpoint\\runtime / model service};
\node[block,below=9mm of endpoint] (remote) {Independent execution\\no live Mach2 monitor};
\node[block,left=of remote] (results) {Returned result bundle\\artifacts / evidence};
\node[block,left=of results] (import) {Mach2 validation gate\\rules / tests / MIK};
\node[block,left=of import] (workspace) {Project workspace\\accepted outputs};
\draw[macharrow] (task) -- (eigen);
\draw[macharrow] (eigen) -- (package);
\draw[macharrow] (package) -- (endpoint);
\draw[macharrow] (endpoint) -- (remote);
\draw[macharrow] (remote) -- (results);
\draw[macharrow] (results) -- (import);
\draw[macharrow] (import) -- (workspace);
\draw[feedback] (workspace.north) .. controls +(0,0.4) and +(0,-0.4) .. (task.south);
\end{tikzpicture}
\caption{Mach2 agent lifecycle with rule checkpoints. Local execution and package construction use instrumented checks; an independently executed package is validated on return before its results are accepted into the project.}
\label{fig:arch}
\end{figure}
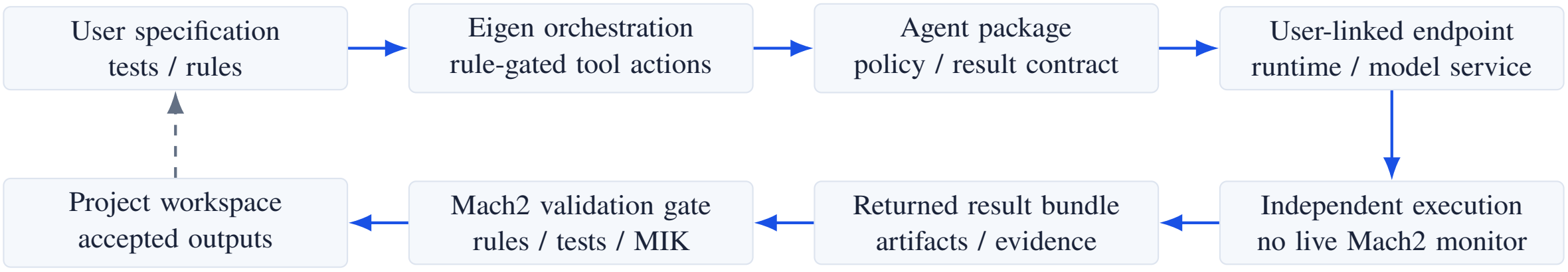

Mach2 can dispatch a generated agent package through a user-linked endpoint when the required service is outside the active workspace or observation boundary. Just-in-time (JIT) binding resolves compatible execution services at invocation time. Figure~\ref{fig:arch} summarizes the independent execution path and the return-validation boundary.

A return bundle records the originating task, package and input versions, endpoint-reported status, output artifacts, and test evidence. At import, Mach2 checks schema conformity, required outputs, task identity, and available domain verification. Failures block acceptance or trigger review. The imported record distinguishes assertions made by the remote executor from results supported by independently executed checks. For example, a remotely computed geometry or simulation result may be syntactically valid yet still fail dimensional, conservation, or convergence requirements.

\section{Comparison with Existing Agent Harnesses}
OpenCode provides an extensible agent runtime and interfaces; DeepSeek Harness provides model-authored workflow composition. Eigen supplies these orchestration capabilities together with engineering tools and a rule-gated task lifecycle. Mach2 can also create agent packages for independent endpoint execution and subsequent result validation. Table~\ref{tab:compare} compares configured mechanisms, rather than asserting that other extensible systems cannot implement analogous checks.

\begin{table}[H]
\caption{Selected execution and policy abstractions across agent harnesses.}
\label{tab:compare}
\centering\rmfamily\fontsize{8.4}{11.1}\selectfont
\renewcommand{\arraystretch}{1.20}
\setlength{\tabcolsep}{4.5pt}
\begin{tabularx}{\linewidth}{@{}p{1.19in}>{\raggedright\arraybackslash}X>{\raggedright\arraybackslash}X>{\raggedright\arraybackslash}X@{}}
\toprule
\textbf{Dimension} & \textbf{OpenCode} & \textbf{DeepSeek Harness} & \textbf{Mach2}\\
\midrule
Orchestration & Configurable agent and tool execution & Model-authored subagent workflows & Eigen-authored engineering workflows\\
Engineering tools & MCP and tool extensibility & Plugin and service architecture & CodeGrapher and CADGrapher\\
Agent generation & User-defined agents and extensions & Agent profiles and workflow bundles & Eigen creates specialized agents and exportable packages\\
Rule governance & Configurable tool permissions & Workspace and sandbox policies & Domain checks with gated acceptance\\
External execution & Application server / SDK & Headless and SDK execution & User-linked agent-package execution\\
Result integration & Application APIs and integrations & Session and service interfaces & Task-linked rule and evidence checks\\
Workspace & Configurable filesystem policies & Workspace and sandbox controls & Launch-selected container project mount\\
\bottomrule
\end{tabularx}
\end{table}

An illustrative task combines a source-code modification, CAD generation, and a calculation at an external solver. CodeGrapher provides AST-grounded change context; user-defined tests and tool restrictions govern local revisions. The exported agent specifies solver inputs and acceptance criteria, while post-return rules check data structure and available physical consistency before the output is incorporated into subsequent work.

\section{Assurance, Knowledge Provenance, and Evaluation}
\subsection{Model-internal knowledge and evidence classification}
Mach2 identifies unsupported model assertions through explicit provenance checks. Claims without a recorded source, AST-backed relation, user specification, or executed verification are flagged as potentially dependent on model-internal knowledge (MIK). This is an \emph{evidence flag}, not direct observation of the model's internal reasoning or a determination that the assertion is false. Such claims can be routed to source retrieval, structured code inspection, TDD, or human assessment. The claim, provenance status, and subsequent verification outcome remain attached to the task record and imported artifacts, consistent with traceable-source validation methods~\cite{sharma2026falseinformation,sharma2024knowledge}.

\subsection{Limits of protocol}
Deterministic checks enforce only the constraints encoded by their predicates and only where calls or artifacts pass through the corresponding validation points. Tests may miss physical errors, an AST cannot prove runtime correctness, and package-level requirements do not establish compliance at an inaccessible endpoint. Evaluation should therefore include deliberate rule violations, untrusted tool outputs, missing provenance, malformed return packages, and adversarially chosen code changes. Existing software benchmarks such as SWE-bench~\cite{swebench} require extension to test numerical and CAD-specific properties

% Controlled comparisons should match model services, tools, engineering tasks, and resource budgets across OpenCode, DeepSeek Harness, supervised Mach2, and detached Mach2 execution. Primary outcomes include the \emph{blocked-violation rate}, \emph{false acceptance rate}, and proportion of artifacts passing independent engineering checks. Other measures include false rejections, rule-evaluation overhead, external integration effort, provenance completeness, and successful resumption after remote execution. Existing software benchmarks such as SWE-bench~\cite{swebench} require extension to test numerical and CAD-specific properties.

\section{Conclusion}
Mach2 combines model-authored orchestration with deterministic checks at defined action, branch-integration, and artifact-acceptance boundaries. Proposed agent changes remain isolated until mandatory checks and integration tests permit commitment to the accepted project state. User-specified TDD, AST-grounded inspection, and history-guided optimization support these engineering decisions, while generated agent packages extend execution to compatible external endpoints. Returned results are reconciled against task contracts and provenance requirements. The assurance offered by this architecture is measurable compliance with specified checks, an important step towards establishing engineering correctness of such AI systems.

Mach2 establishes an extensible foundation for subsequent generations of scientific agentic systems, emphasizing privacy-preserving computation, model governance, and low-latency execution. Future iterations of the Mach series, including Mach3, will extend this architecture through runtime classifiers for identifying potentially invalid or policy-inconsistent tool invocations, coupled with deterministic enforcement mechanisms. Further developments will address adaptive orchestration of complex scientific workflows, building upon scientific reasoning frameworks, such as Ref.~\cite{sharma2026thesis}, as well as multimodal tool integration and connectivity with persistent user conversational workspaces. These capabilities will enable controlled context transfer across agents, sessions, and execution environments while maintaining execution traceability, knowledge provenance, and consistent governance policies.

\section*{Acknowledgements} This work was supported by the Center for Prediction, Reasoning, and Intelligence for Multiphysics Exploration (C-PRIME), a PSAAP-IV project funded by the U.S. Department of Energy under grant DE-NA0004264 (program manager: Dr. David Etim).

%%%%%%%%%%%%%%%%%%%% Bib

\bibliographystyle{unsrtnat}
\nocite{*}
\bibliography{references}
\end{document}